\documentclass[a4paper,12pt]{article}
\usepackage{a4p}
\usepackage{pennames}
\usepackage{graphics}

\newcommand{\Mnutau}{\ensuremath{m_{\mathrm{\nu_{\tau}}}}}
\newcommand{\TheRawNumber}{39.6}
\newcommand{\TheFinalNumber}{43.2}
\newcommand{\TheCombinedNumber}{27.6}
\newcommand{\mylumi}{155}
\newcommand{\Nselected}{22}
\newcommand{\Bdqq}{\ensuremath{\mathrm{e^+e^-}\rightarrow \mathrm{q\bar q}}}
\newcommand{\Bqq}{\ensuremath{\mathrm{q\bar q}}}
\newcommand{\Bpizero}{\ensuremath{\mathrm{3\Pgp^{\pm} \Pgp^0}}}

\newcommand{\BKKp}{\ensuremath{\mathrm{K^0_SK^0_S\Pgp^{\pm}}}}
\newcommand{\Bttff}{\ensuremath{\mathrm{\tau^+\tau^- f\bar f}}}
\newcommand{\Dalitz}{\ensuremath{\mathrm{\Pgp^0\rightarrow e^+e^-\Pgg}}}
\newcommand{\Bdpizero}{\ensuremath{\mathrm{\tau \rightarrow 3\Pgp^{\pm} \Pgp^0\Pgngt}}}

\newcommand{\Bdthreepi}{\ensuremath{\mathrm{\tau \rightarrow 3\Pgp^{\pm} \Pgngt}}}
\newcommand{\Bthreepi}{\ensuremath{\mathrm{3 \Pgp^{\pm}}}}
\newcommand{\BdKKp}{\ensuremath{\mathrm{\tau \rightarrow K^0_SK^0_S\Pgp^{\pm} \Pgngt}}}
\newcommand{\Bdfivepizero}{\ensuremath{\mathrm{\tau \rightarrow 5\Pgp^{\pm} \Pgp^0 \Pgngt}}}
\newcommand{\Bdttff}{\ensuremath{\mathrm{e^+e^-}\rightarrow \tau^+\tau^-\mathrm{f\bar f}}}
\newcommand{\Sfivepi}{\ensuremath{\mathrm{5\Pgp}}}

\newcommand{\Sdfivepi}{\ensuremath{\mathrm{\Pgt \rightarrow 5\Pgp^{\pm} \Pgngt}}}
\newcommand{\Makotodecay}{\ensuremath{\mathrm{\Pgt^+ \Pgt^- \rightarrow 3h^{\pm} \bar \Pgngt + 3h^{\mp} \Pgngt}}}
\newcommand{\smaller}{\ensuremath{\mathrm{<}}}
\newcommand{\Stau}{\ensuremath{\mathrm{\tau}}}
\newcommand{\epem}{\ensuremath{\mathrm{e^+e^-}}}
\newcommand{\mpmm}{\ensuremath{\mathrm{\mu^+\mu^-}}}
\newcommand{\Minv}{\ensuremath{m_{\mathrm{5\pi}}}}
\newcommand{\Efivepi}{\ensuremath{E_{\mathrm{5\pi}}}}
\newcommand{\dEdx}{\ensuremath{\mathrm{d}E/\mathrm{d}x}}

\newcommand{\MyEffi }{$\mathrm{ (9.3 \pm 0.6)\%}$}

\newcommand{\ra}{\rightarrow}

\newcommand{\Pizero}{\ensuremath{\mathrm{\Pgp ^0}}}
\newcommand{\Pgtm}{\ensuremath{\mathrm{\tau^-}}}
\newcommand{\Pgtp}{\ensuremath{\mathrm{\tau^+}}}

\newcommand{\rb}[1]{\raisebox{1.5ex}[-1.5ex]{#1}}

\newcommand{\eett}{\ensuremath{\Pep\Pem\ra\Pgtp\Pgtm}}
\newcommand{\eeqq}{\ensuremath{\Pep\Pem\ra \mathrm{q \bar q}}}
\newcommand{\eettff}{\ensuremath{\Pep\Pem\ra\Pgtp\Pgtm \mathrm{f \bar f}}}

\newcommand{\GeV}{\ensuremath{\rm GeV}}
\newcommand{\MeV}{\ensuremath{\rm MeV}}
\newcommand{\ipb}{\ensuremath{\rm pb ^{-1}}}
\newcommand{\micron}{\ensuremath{\rm \mu m}}

\begin{document}
\begin{titlepage}
\begin{center}{\large\bf   EUROPEAN LABORATORY FOR PARTICLE PHYSICS
}\end{center}\bigskip
\begin{flushright}
       CERN-EP/98-055   \\ April 15, 1998 \\ Journal Version \\ June 5, 1998
\end{flushright}
\bigskip\bigskip\bigskip\bigskip\bigskip
\begin{center}{\huge\bf\boldmath An upper limit for the \Stau --neutrino mass \\ from \Sdfivepi\ decays 
}\end{center}\bigskip\bigskip
\begin{center}{\LARGE The OPAL Collaboration
}\end{center}\bigskip\bigskip
\bigskip\bigskip\bigskip
\bigskip\bigskip\bigskip
\bigskip\bigskip\bigskip
\bigskip\bigskip\bigskip
\bigskip\begin{center}{\large\bf  Abstract}\end{center}
\noindent
An upper limit for the \Stau --neutrino mass has been determined
from the decay \Sdfivepi\ using data collected
with the OPAL detector from 1991 to 1995 in \epem\,collisions
at $ \sqrt{s} \approx M_\mathrm{Z}$. 
A limit of \TheFinalNumber\ \MeV\ 
at 95\% CL is
obtained using a two--dimensional method
in the \Sfivepi~invariant mass and 
energy distribution from \Nselected\ selected events.
Combining this result with OPAL's previously published
measurement using \Makotodecay \, decays, a new combined limit of
\Mnutau\smaller\,\TheCombinedNumber\,\MeV\ (95\% CL) is obtained.
\bigskip\bigskip\bigskip
\bigskip\bigskip\bigskip
\bigskip\bigskip\bigskip
\bigskip\bigskip\bigskip
\bigskip\bigskip\bigskip
\begin{center}{\large
(Submitted to European Physical Journal C) 
}\end{center}
\end{titlepage}
\begin{center}{\Large        The OPAL Collaboration
}\end{center}\bigskip
\begin{center}{
K.\thinspace Ackerstaff$^{  8}$,
G.\thinspace Alexander$^{ 23}$,
J.\thinspace Allison$^{ 16}$,
N.\thinspace Altekamp$^{  5}$,
K.J.\thinspace Anderson$^{  9}$,
S.\thinspace Anderson$^{ 12}$,
S.\thinspace Arcelli$^{  2}$,
S.\thinspace Asai$^{ 24}$,
S.F.\thinspace Ashby$^{  1}$,
D.\thinspace Axen$^{ 29}$
% ...
}\end{center}\bigskip
\bigskip
$^{  1}$School of Physics and Astronomy, University of Birmingham,
Birmingham B15 2TT, UK
\newline
$^{  2}$Dipartimento di Fisica dell' Universit{\`a} di Bologna and INFN,
I-40126 Bologna, Italy
\newline
$^{  3}$Physikalisches Institut, Universit{\"a}t Bonn,
D-53115 Bonn, Germany
\newline
$^{  4}$Department of Physics, University of California,
Riverside CA 92521, USA
\newline
$^{  5}$Cavendish Laboratory, Cambridge CB3 0HE, UK
\newline
$^{  6}$Ottawa-Carleton Institute for Physics,
Department of Physics, Carleton University,
Ottawa, Ontario K1S 5B6, Canada
\newline
$^{  7}$Centre for Research in Particle Physics,
Carleton University, Ottawa, Ontario K1S 5B6, Canada
\newline
$^{  8}$CERN, European Organisation for Particle Physics,
CH-1211 Geneva 23, Switzerland
\newline
$^{  9}$Enrico Fermi Institute and Department of Physics,
University of Chicago, Chicago IL 60637, USA
\newline
$^{ 10}$Fakult{\"a}t f{\"u}r Physik, Albert Ludwigs Universit{\"a}t,
D-79104 Freiburg, Germany
\newline
$^{ 11}$Physikalisches Institut, Universit{\"a}t
Heidelberg, D-69120 Heidelberg, Germany
\newline
$^{ 12}$Indiana University, Department of Physics,
Swain Hall West 117, Bloomington IN 47405, USA
\newline
$^{ 13}$Queen Mary and Westfield College, University of London,
London E1 4NS, UK
\newline
$^{ 14}$Technische Hochschule Aachen, III Physikalisches Institut,
Sommerfeldstrasse 26-28, D-52056 Aachen, Germany
\newline
$^{ 15}$University College London, London WC1E 6BT, UK
\newline
$^{ 16}$Department of Physics, Schuster Laboratory, The University,
Manchester M13 9PL, UK
\newline
$^{ 17}$Department of Physics, University of Maryland,
College Park, MD 20742, USA
\newline
$^{ 18}$Laboratoire de Physique Nucl{\'e}aire, Universit{\'e} de Montr{\'e}al,
Montr{\'e}al, Quebec H3C 3J7, Canada
\newline
$^{ 19}$University of Oregon, Department of Physics, Eugene
OR 97403, USA
\newline
$^{ 20}$Rutherford Appleton Laboratory, Chilton,
Didcot, Oxfordshire OX11 0QX, UK
\newline
$^{ 22}$Department of Physics, Technion-Israel Institute of
Technology, Haifa 32000, Israel
\newline
$^{ 23}$Department of Physics and Astronomy, Tel Aviv University,
Tel Aviv 69978, Israel
\newline
$^{ 24}$International Centre for Elementary Particle Physics and
Department of Physics, University of Tokyo, Tokyo 113, and
Kobe University, Kobe 657, Japan
\newline
$^{ 25}$Institute of Physical and Environmental Sciences,
Brunel University, Uxbridge, Middlesex UB8 3PH, UK
\newline
$^{ 26}$Particle Physics Department, Weizmann Institute of Science,
Rehovot 76100, Israel
\newline
$^{ 27}$Universit{\"a}t Hamburg/DESY, II Institut f{\"u}r Experimental
Physik, Notkestrasse 85, D-22607 Hamburg, Germany
\newline
$^{ 28}$University of Victoria, Department of Physics, P O Box 3055,
Victoria BC V8W 3P6, Canada
\newline
$^{ 29}$University of British Columbia, Department of Physics,
Vancouver BC V6T 1Z1, Canada
\newline
$^{ 30}$University of Alberta,  Department of Physics,
Edmonton AB T6G 2J1, Canada
\newline
$^{ 31}$Duke University, Dept of Physics,
Durham, NC 27708-0305, USA
\newline
$^{ 32}$Research Institute for Particle and Nuclear Physics,
H-1525 Budapest, P O  Box 49, Hungary
\newline
$^{ 33}$Institute of Nuclear Research,
H-4001 Debrecen, P O  Box 51, Hungary
\newline
$^{ 34}$Ludwigs-Maximilians-Universit{\"a}t M{\"u}nchen,
Sektion Physik, Am Coulombwall 1, D-85748 Garching, Germany
\newline
\bigskip\newline
$^{  a}$ and at TRIUMF, Vancouver, Canada V6T 2A3
\newline
$^{  b}$ and Royal Society University Research Fellow
\newline
$^{  c}$ and Institute of Nuclear Research, Debrecen, Hungary
\newline
$^{  d}$ and Department of Experimental Physics, Lajos Kossuth
University, Debrecen, Hungary
\newline
$^{  e}$ on leave of absence from the University of Freiburg
\newline
\newpage
\section{ Introduction}
\label{Introduction}
The question of whether neutrinos
have
mass is one of
the outstanding 
issues
in particle physics, astrophysics,
and cosmology.
Massive neutrinos are strong candidates
for solving the dark matter problem of
the universe\,\cite{theory:darkmatter}. 
Of the three neutrino species, the \Stau --neutrino,
is likely to have the largest mass.
For instance, in the `see--saw'
mechanism\,\cite{Theory:seesaw}
a mass hierarchy exists between neutrinos and their corresponding 
lepton partners, rendering the \Stau --neutrino the 
heaviest of the three known neutrino types.  

On the basis of cosmological arguments a stable \Stau --neutrino
with a mass
larger than a few $ \mathrm{eV}$  cannot exist\,\cite{Theory:gravlimit},
however unstable neutrinos may be more massive\,\cite{Theory:bigbang}.
Previously, 
OPAL has published an upper limit
on \Mnutau \,of 74\,\MeV \ based on
one event
in the 
rare  \Sdfivepi\ decay channel\,\cite{Opal:first} in the 1992 data.
In this final state the
distribution of events in energy \Efivepi\ and invariant mass \Minv\
of the hadronic system
at the two--dimensional limit of the kinematic range is sensitive to \Mnutau .  

For this paper, all data collected by OPAL from 1991 to 1995
have been analysed to obtain a new limit on the tau--neutrino mass
using again the \Sdfivepi\ decay channel.
Compared to the previous analysis,
 the number of events considered has increased fivefold. 
\section{The OPAL detector and simulation}
\label{Opaldetector}
A detailed description of the OPAL detector can be found
in\,\cite{Opal:detector1}.
Subdetectors which are particularly
relevant to the present analysis are briefly described below.

The
central detector consists of a set of tracking chambers providing
charged particle tracking over 96\% of the
solid angle inside a
0.435\,T uniform magnetic field parallel to the beam axis.
Starting with the
innermost components, it consists of a high precision silicon
microvertex detector, a precision vertex
drift chamber, a large volume jet chamber
and a set of $z$--chambers\footnote{
The OPAL coordinate system is defined so that $z$ is the 
coordinate parallel to the beam axis, the radius $r$ is the
coordinate normal
to the beam axis, $\mathrm{\phi}$ is the azimuthal angle and 
$\mathrm{\theta}$ is the polar angle with respect to $z$.}
measuring the track coordinate
along the beam direction.

From 1991 onwards a silicon strip microvertex detector was
also present, consisting of 
two concentric layers
with readout strips at 50\,\micron\ pitch,  oriented
for azimuthal ($\mathrm{\phi}$) coordinate measurement\,\cite{Opal:Silicon1}.
In 1993 a new silicon strip microvertex detector with
$z$--coordinate readout in addition 
was installed\,\cite{Opal:Silicon2}.

The jet chamber is designed to combine good space and double
track resolution\,\cite{Opal:cjperformance}, which is important for this analysis.
It consists of 159 layers of axial
anode wires, which are located between radii of 255\,mm and 1835\,mm.
The efficiency for separating hits from two
adjacent particles in the jet chamber is approximately 80\% for
distances between two hits of 2.5\,mm in the projection
on the $r$--$\phi$ plane\,\cite{Opal:cjperformance} and
drops rapidly for smaller hit distances. 
The transverse momentum resolution
of isolated tracks is 
$\sigma_{p_t}/p_t = \sqrt{(0.02)^2 + (0.0015 \cdot p_t \,[\GeV ])^2}$.
The jet chamber also provides energy loss measurements for particle
identification (\dEdx ). The \dEdx\ resolution is
 $\frac{\sigma(\mathrm{d}E/\mathrm{d}x)}{\mathrm{d}E/\mathrm{d}x} = 3.2 \%$ 
for minimum ionizing pions in jets with the maximum number of hits (159), 
resulting in a $\mathrm{\pi}$--e separation of at least 2
standard deviations up to 
momenta of 14\,\GeV\,\cite{Opal:hausch1,Opal:hausch2}.

A lead-glass electromagnetic calorimeter (ECAL) located outside the
magnet coil covers the full azimuthal range with excellent hermeticity
in the polar angle range of $|\cos \theta |<0.82$ for the barrel
region and $0.81<|\cos \theta |<0.98 $ for the endcap region.

The Monte Carlo samples used in this analysis consist of 
1.5 million \eett\,\cite{Generator:tauola-24,Generator:koralz-40},
8.5 million
\eeqq\,\cite{Generator:jetset74} and 10\thinspace 500 \eettff\,\cite{Generator:fermisv}
 events, which are processed through
the OPAL detector simulation\,\cite{Opal:gopal}. 
These samples correspond to about
8, 2 and 20 times the data luminosity, respectively.

\section{Event selection}
\label{Selection}
Data collected during the years 1991 to 1995,
corresponding to an integrated luminosity of \mylumi\,\ipb\, 
and almost 200\thinspace 000 recorded
e$^+$e$^- \rightarrow \tau^+ \tau^-$ events 
have been analysed.
The event selection
is performed in two steps.
First, the preselection selects \Stau\ candidates with five charged 
tracks in a cone. In the second
step,
background \Stau\ decays and remaining non--\Stau\ events are rejected.
\subsection{Preselection}
\label{Preselection}

A cone jet algorithm\,\cite{Opal:cone} is employed to
assign all tracks
and electromagnetic clusters to cones with a half opening angle of
$\mathrm{35^\circ}$. 
For each event exactly two cones are required.
The `signal' cone is required to contain exactly five charged tracks
with unit total charge.
The other `recoil' cone is required to contain at least one track.

All tracks are required to satisfy
the following conditions:
\begin{itemize}
\item $p_t >$ 100\,\MeV, where $p_t$ is the momentum 
component transverse to the beam direction;
\item at least 20 hits in the central jet chamber.
This restricts the acceptance of the detector to tracks
with $\mathrm{|\cos\theta|~<~0.963}$;

\item the distance $\vert d_0 \vert$ of closest approach of the track to the 
beam axis must be smaller than 2\,cm. The displacement of 
the track along the beam axis from the nominal interaction 
point at the point of closest approach to the beam must be
less than 75\,cm;
\item  the radial distance from the beam axis of the first 
hit in the jet chamber associated to a track
must be smaller than 120\,cm.
\end{itemize}

To reject non--\Stau \ events the OPAL standard selection of
\Stau\ pairs is adopted\,\cite{Opal:h3paper}.
The multihadronic background (\Bdqq) is 
reduced by demanding a maximum of six 
tracks and 10 electromagnetic clusters in the event.
A cluster is defined as a group of contiguous lead-glass
blocks which has a minimum energy of 100\,\MeV\ in the barrel 
or 200\,\MeV\ in the endcap. 
The requirement on the maximum number of tracks
leads to a 5--1 topology
of tracks in the signal and recoil cone for all preselected
events.

\subsection{Final selection}
\label{eventselection}
\begin{figure}[htb]
\begin{center}
\resizebox{17cm}{!}{\includegraphics{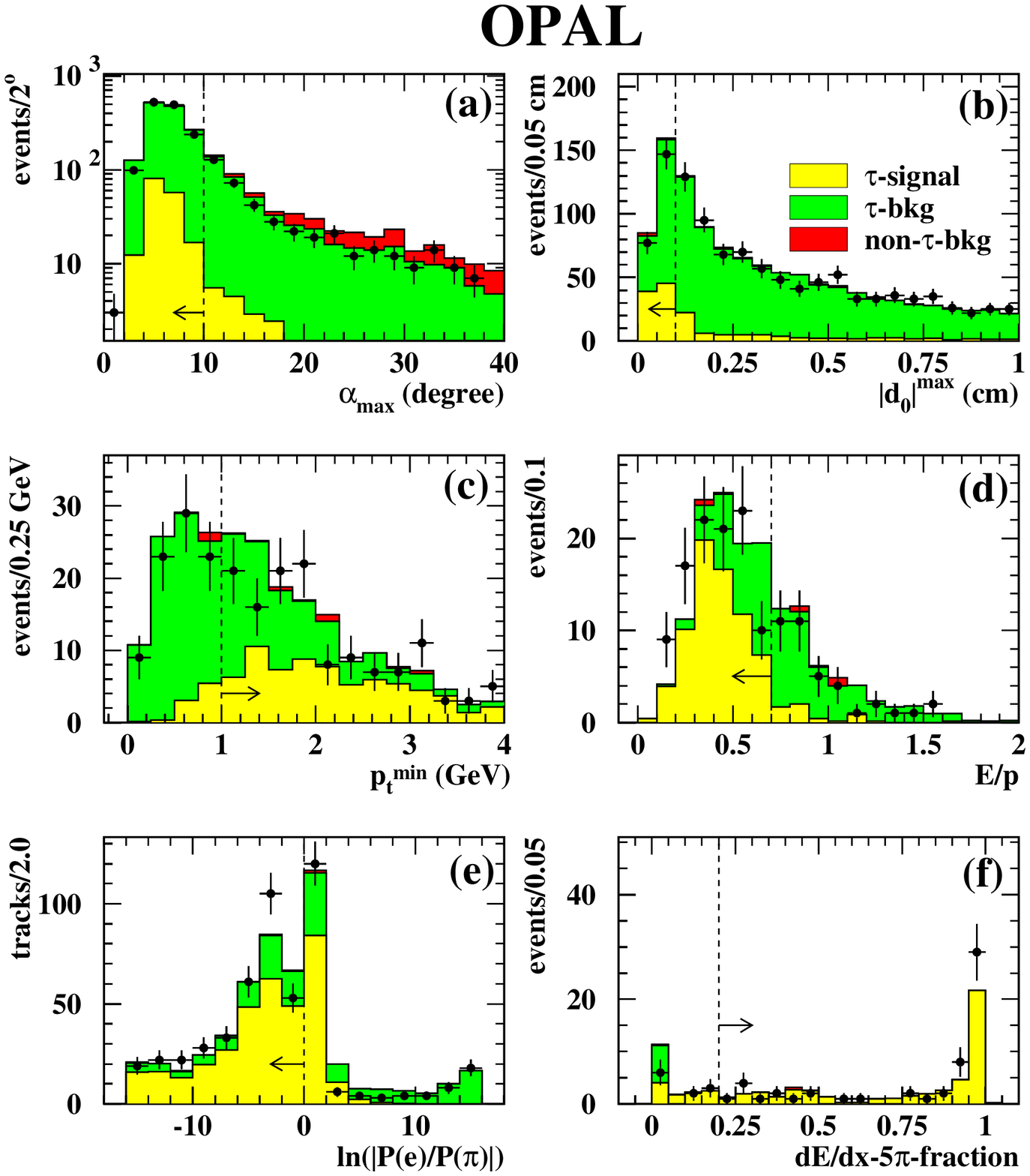}}
\end{center}
\caption{\it Distributions of the most important quantities used 
for background suppression in the selection. 
The points with error bars are the data.
The histograms denote the Monte Carlo expectation,
normalized to the luminosity of the data.
The cut order and the cut definitions are 
described in section\,\ref{eventselection}.
All previous cuts have been applied in each plot.
The dashed lines indicate the positions of cuts and the
arrows point into the selected region.}
\label{fig:break}
\end{figure}

The background from other \Stau\ decays and from multihadronic events
is reduced by rejecting events if the maximum opening
angle $\mathrm{ \alpha_{max}}$ between two tracks in
the signal cone is larger than $\mathrm{10^\circ}$
(see figure\,\ref{fig:break}a).
  
The remaining background is dominated by \Stau\ decays
into three charged particles accompanied 
by a photon conversion to an \epem\ pair thus creating 
a final state with five charged tracks. 
To reject these events
the following cuts are applied on the signal side.
Events where any track has an impact parameter $ \vert d_0 \vert$
with respect to the beam axis larger
than 0.1\,cm are rejected (figure\,\ref{fig:break}b).
The minimum transverse momentum $p_{t}^{\mathrm{min}}$
of any track has to be larger
than 1\,\GeV\ (figure\,\ref{fig:break}c).
Furthermore  the fraction $ E/p $ is required to 
be smaller
than 0.7 (figure\,\ref{fig:break}d),
where  $ E $  is the deposited energy in the electromagnetic
calorimeter and $ p $ is the sum
of the momenta of the five charged
tracks.

The next two selection cuts exploit the \dEdx \ information
of the jet chamber together with the information from
the silicon microvertex detector
mainly to reject events
from \Bdpizero \ 
where a photon from the \Pizero\,decay 
has converted.
If a track appears to be more likely to originate
from an electron than
from a pion ($P_e>P_{\pi}$) or 
if insufficient \dEdx \ 
information is 
available, at least one associated hit in the silicon
microvertex detector
is required.
Here $ P_{ \pi}  (P_e) $ is the
 $ \mathrm{\chi ^2} $-probability
 that the track is consistent with the pion (electron) hypothesis
derived from the \dEdx \ and momentum measurement
(figure\,\ref{fig:break}e).
A test on the total likelihood for
 the 5\Pgp \ final state is also performed.
The fraction $ P(5 \pi)/(P(5 \pi )+\sum P(3 \pi) P(\mathrm{e^+e^-}))$
must favour the $\mathrm{ 5\pi}$ hypothesis ($>$\,0.2), where
$P( 5 \pi ) = \prod_{i=1}^{5} P_{ \pi}( i ) $
and $\sum P( 3 \pi) P(\mathrm{e^+e^-}) $ is the sum of the 
combinatorial possibilities of 
three particles to be pions
 and two to be oppositely charged
electrons (figure\,\ref{fig:break}f).

Events with a cluster in the electromagnetic calorimeter
within the signal cone
with 
energy
more than 4\,\GeV\ and not associated with a track
are discarded because this signature probably comes from
a photon.

Additional quality cuts on the tracks have been applied to
 ensure 
a good reconstruction of the \Sfivepi \ system.
Each track is required to have at least 40 hits in the central 
jet chamber.
Each track fit
must have a $\mathrm{\chi^2}$ per degree of freedom smaller than 2.
Given the 
high density of tracks in the \Sfivepi\ final state,
this cut aims to reject 
events with falsely reconstructed tracks 
due to spatial distortions of the chamber hits
or due to hit misassignments by the pattern recognition algorithm.
Furthermore events are rejected where the angle between a
high--momentum track ($p_t > $ 15\,\GeV) and any wire plane of the 
jet chamber is smaller than 0.3$^\circ$.
This cut
eliminates events with tracks which may be badly reconstructed due to
distortions of the drift field in direct proximity of the anode and
cathode planes.

After this selection \Nselected\ candidate events remain.
The positions
of these events in the \Efivepi -- \Minv\ plane are shown 
in figure\,\ref{fig:allselected}.
According to the Monte Carlo (MC) simulation, 
the selection efficiency after all cuts 
is \MyEffi, where the error is statistical only.
\section{Background}
\label{background}
A high purity data sample is required for an unbiased
neutrino mass limit.
The background can be divided into two classes: (a) \Stau --pair events
with a decay misidentified as \Sdfivepi\ on the signal side and (b)
non-\Stau \ events with a topology similar to \Stau \ decays. 
The reconstructed mass and hadronic
energy of these events may be accidentally located close to the
kinematic boundary in the \Efivepi --\Minv\ plane,
leading to an artificially low neutrino mass limit. 

The background is
estimated from
Monte Carlo event samples described in section\,\ref{Opaldetector}.
For the \Stau \ background class (a) we have considered the following
decay channels:
\begin{itemize}
\item {\bf \Bdthreepi : } The \Stau\ decays into three
charged pions one of which undergoes a hadronic interaction
within the beam pipe or the vertex detector. The final state consists of
five pions tending to higher invariant masses.
\item {\bf \Bdpizero : } 
One of the photons from the \Pizero\,decay
converts in the detector material 
or a Dalitz decay (\Dalitz ) occurs.
If the two electrons are misidentified
as pions, the reconstructed invariant mass is artificially high.
\item {\bf \BdKKp :}
If both the $ \mathrm{ K^0_S } $ decay very close to the
interaction point,
the final state \Sfivepi\ system cannot be distinguished from
the signal. The expected bias is small, because the 
mass hypothesis for all tracks is the same as for \Sdfivepi.
\item {\bf \Bdfivepizero : } 
A 20\% contamination from these events is expected in the data. 
As explained in
section\,\ref{systematics} these events cannot bias the measurement
to lower mass limits and
this is therefore not a serious background.
\end{itemize}

In this background class only one \Bdpizero\ MC event
passes the selection corresponding to 0.11 data events in the full
\Efivepi --\Minv\ region.
The fraction of \Mnutau --sensitive or `effective' \Stau\ background
is smaller. An event is denoted as \Mnutau --sensitive, if its position
in the \Efivepi --\Minv\ plane could lead to a mass limit of below
100\,\MeV.
Based on the \Efivepi --\Minv\ distribution
of MC events in the observed background decay channel it is estimated
that less than one tenth of these background events would
influence the neutrino mass limit.

\begin{table}[htb]
\begin{center}
\begin{tabular}{|c|c|c|c|}
\hline
Background & Background & Expected number & Effective number\\
class      & source     &    of events    &  of events      \\
\hline
 & \Bpizero   & $ \mathrm{ 0.11 \pm 0.11} $  &  \\
 \rb{(a) \Pgt $ \ra$ X} & \BKKp , \Bthreepi  & $ \mathrm{ < 0.14 (68 \% CL)} $ & \rb{ $ 0.01 \pm 0.01 $ } \\
\hline
                   & \Bqq    & $ \mathrm{ 0.45 \pm 0.45 } $ & \\
\rb{(b) non--\Pgt} & \Bttff  & $ \mathrm{ < 0.06 (68\% CL) } $ & \rb{ $ \mathrm{ 0.04 \pm 0.04} $ } \\
\hline 
\hline 
\multicolumn{2}{|c|}{total} &  $ \mathrm{ 0.56 \pm 0.49 } $ & $ \mathrm{ 0.05 \pm 0.04}$ \\
\hline 
\end{tabular} 
\caption{ \it Expected background in the selected sample}
\label{tab:back}
\end{center}
\end{table}

Out of the multihadronic MC samples, background class (b), one event
is selected. 
It has many
clusters 
in the electromagnetic calorimeter
and the reconstructed mass (2.3 \GeV ) of the signal cone is 
too high to be compatible with a \Stau \ decay.
Normalized to the data luminosity this event corresponds
to an expected $ \mathrm{q \bar q}$ background of 0.45 events.
For the estimation of the effective \Bqq\ background the multiplicity
cuts for tracks on the recoil cone and for clusters in the event are
relaxed. 
Then 11 \Bqq\ MC events are selected.
Three of these are located inside the kinematically
allowed signal region and only one event lies close enough
to the boundary such that its consideration would have an
impact on the extracted limit.
It is therefore concluded that 
the
\Bqq\ background that could affect \Mnutau\ is
only about 0.04 events.

The expected multihadron background has also been cross--checked
using data events looking for the 5--2 event topology
after relaxing the corresponding multiplicity cut.
One such event in the data sample is observed 
with 1.3 expected from the \Stau \ MC,
confirming the direct MC prediction of the $\mathrm{q\bar q} $
background.

The background from four fermion events (\Bdttff) originates 
mainly from a \Stau \ decay into three charged tracks
combined with the 
fermion--antifermion pair.
In order to estimate the contribution of this background,
Monte Carlo samples are 
used where the $ \mathrm{ f \bar f}$ pair is either a 
$\mathrm{q\bar q}$, \epem , or \mpmm pair.
No such event passes the selection.
 
The total effective background is therefore expected to be about 0.05
events and is considered as negligible.
The estimated background is summarized in table\,\ref{tab:back}.

\section{Determination of the mass limit}
\label{method}
\subsection{Likelihood analysis}
\label{likelihood}
The upper limit on the \Stau --neutrino mass is obtained employing 
a likelihood analysis.
The probability $ P_{i}(m_{i},E_{i} \vert\,m_{\nu} )$
for observing each selected event $i$
at the position ($m_i,E_i$) within the kinematic 
plane is derived as function of the neutrino mass
$m_{\nu}$.
To obtain this probability the theoretical prediction
 ${\cal P}(m,E \vert m_{\nu})$
for measuring
the observed distribution in the \Efivepi --\Minv\  plane
is convolved 
with the
experimental resolution $R$ and the detection
\mbox{efficiency $\epsilon$}.
Hence the probability can be written as 
\[
P_{i}(m_{i},E_{i} \vert\,m_{\nu} ) = \frac{\int dm \int dE
{\,\cal P}(m,E \vert\,m_{\nu})
R(m-m_i,E-E_i,\sigma_{m_i},\sigma_{E_i},\rho) \epsilon(m,E)}
{\int dm \int dE
{\,\cal P}(m,E \vert\,m_{\nu})
\epsilon(m,E)} \;\; . 
\label{Faltungsintegral}
\]

The theoretical prediction
${\cal P}(m,E \vert\,m_{\nu})$
is generated as a function
of the neutrino mass \Mnutau\ using
KORALZ--TAUOLA\,\cite{Generator:koralz-40,Generator:tauola-24}
including initial--state 
radiation. The neutrino mass was restricted to positive 
values.
The detection efficiency $\epsilon(m,E)$ is derived from Monte Carlo
using full detector simulation\,\cite{Opal:gopal}.
The function used to describe the experimental resolution 
$R(m-m_i,E-E_i,\sigma_{m_i},\sigma_{E_i},\rho)$
is a two--dimensional Gaussian. 
The corresponding parameters, the errors on the invariant
mass $\sigma_{m_i}$, on the energy $\sigma_{E_i}$, and 
the correlation $\rho$  between them, 
are described in the following section.

The efficiency $\mathrm{\epsilon(m,E) }$ is, to a good approximation,
independent of $m$ and $E$ ($\epsilon = 0.093 \pm 0.014$).
Hence the formula for $ P_{i}(m_{i},E_{i} \vert\,m_{\nu} )$
simplifies to 
\[
P_{i}(m_{i},E_{i}\vert\,m_{\nu}) = \frac{1}{k} \int dm \int dE
{\,\cal P}(m,E \vert\,m_{\nu})
R(m-m_i,E-E_i,\sigma_{m_i},\sigma_{E_i},\rho)   
\label{SimFaltungsintegral}
\]
with a constant $k = \int dm \int dE {\,\cal P}(m,E \vert\,m_{\nu})$.

The sensitivity of the neutrino mass limit to the efficiency
is small (see section\,\ref{systematics} below).
\subsection{Experimental Resolution}
\label{resolution}
Most of the events lie well inside the kinematically allowed region
in figure\,\ref{fig:allselected} 
such that they do not contribute significantly to the mass limit.
An event is denoted `insensitive', if a limit below 100\,\MeV\ 
cannot be achieved using this event alone. 
For these events
the errors on the track parameters are propagated to 
errors on the invariant mass,
energy and the correlation. 
These errors 
are then taken as input for the binormal resolution function.

For the other events which are located near or outside the kinematic
boundary (`sensitive' events),
the exact form of the error ellipse in the \Efivepi --\Minv\ plane
is of crucial importance for the determination of the limit on 
\Mnutau.
Therefore an approach is used which considers the strong
dependence of the resolution $ R $ on the specific topology 
of the event (i.e., hits in particular subdetectors
of the tracking system and
susceptibility to reconstruction errors).

For these events the measured four--momenta of each event are used
as input for
the detector simulation\,\cite{Opal:gopal} 
and reprocessed through the full simulation several thousand times.
The subsample of these events that all
have the same
reconstruction properties
as the original event (e.g., same number of tracks with hits in the
silicon microvertex detector and in the $z$--chambers)
are used to determine the experimental resolution.
This is done by fitting a 
two--dimensional Gaussian function with correlation
in an unbinned likelihood 
fit to the
reconstructed \Efivepi --\Minv\ spectrum
of the simulated events.
A small non--Gaussian (`tail') fraction of the distributions 
is eliminated
to a large extent by discarding events deviating by more than 
3 standard deviations from the fitted mean.
The fraction of discarded events is about 2\%, and
the remaining events
are used to determine 
the parameters of the resolution function.
The typical mass and energy resolutions for
\Sfivepi \  decays in MC are 20--25\,\MeV\ and
500\,\MeV, respectively.

In order to assess a possible bias introduced by the tails
in the \Efivepi --\Minv\  distributions
the influence of the fraction of events residing in the tails
has been estimated.
First, a sum of two 2--dimensional Gaussian functions has
been used in the fit, 
where the second wider Gaussian is introduced to describe the tails.
The fraction of this Gaussian has been varied by
$ \mathrm{ \pm 50 \% } $ to estimate the impact of the tails.
Alternatively, the sum of a Gaussian
and a flat pedestal distribution on 
the \Efivepi --\Minv\ plane has been fitted.
In both approaches the effect on the extracted limit
on \Mnutau\ does not vary
by more than 3.5\,\MeV.

\begin{figure}[htb]
\begin{center}
\resizebox{17cm}{!}{\includegraphics{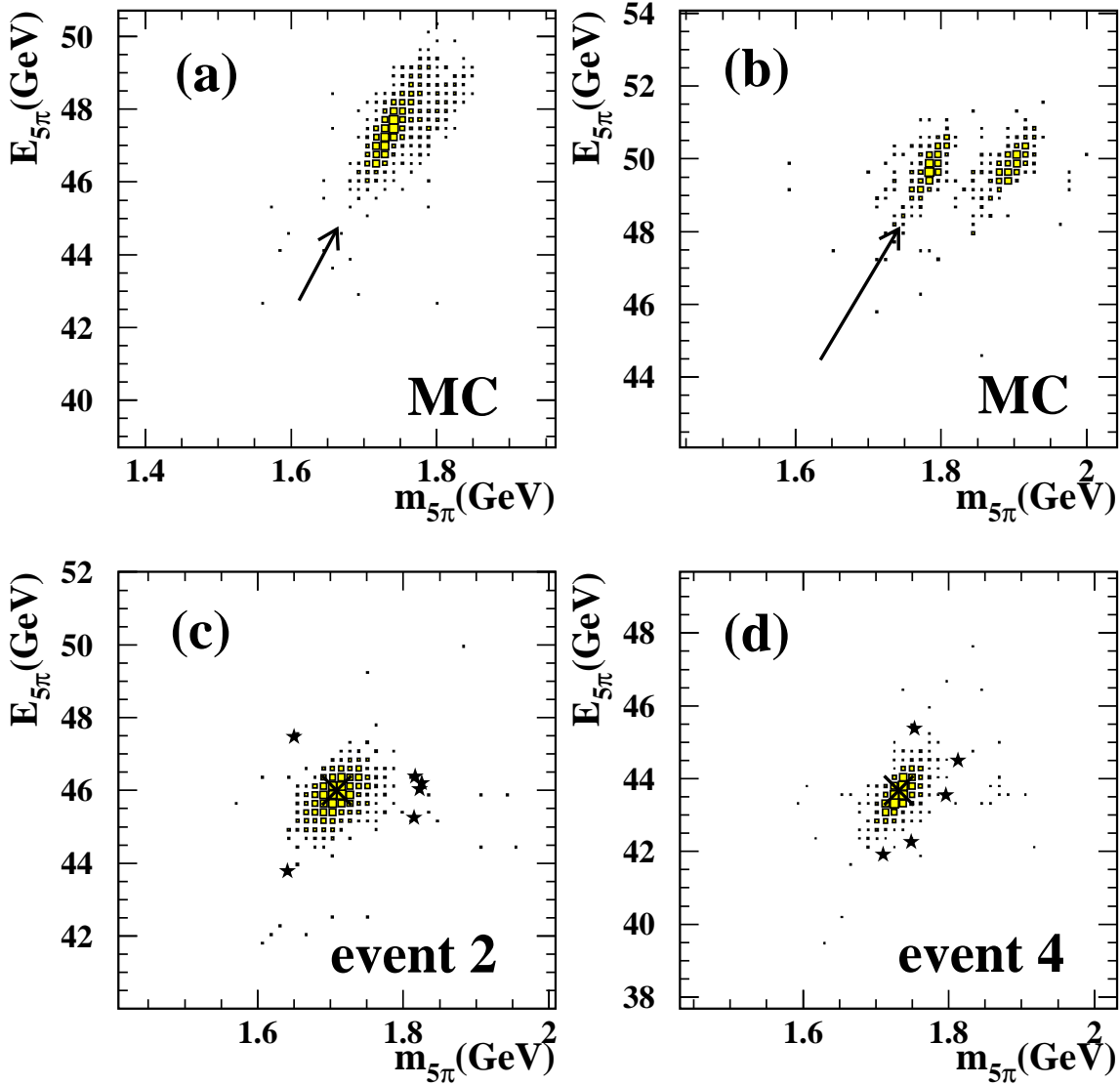}}
\end{center}
\caption{\it
OPAL simulation of events. Figures (a) and (b) show the output
of the simulation procedure with one MC event to each plot.
The starting point of the arrows shows the true position 
in the \Efivepi --\Minv\ plane, the end point the reconstructed 
position. The latter was used as input for the
simulation (section 5.2).
In cases where a large discrepancy between the true and reconstructed
values occurs, the simulation shows significant
defects such as shifts (a)
or ambiguities (b) between simulation input and output.
Figures (c) and (d) show the analogous output for the 
two most sensitive data events. The crosses denote the
input values. The output distributions are centered
around those points and are unambiguous. The stars show
the additional positions where the resolution was determined.
}
\label{fig:tail}
\end{figure}

It is essential for this analysis that all events taken into
consideration for the mass limit are well measured.
Particularly for decays in which two or more tracks cross within the 
drift chamber volume or
approach each other closely,
an incorrect hit assignment may cause biases
in the tails of the \Efivepi --\Minv\ distributions.
Such biases have been studied
using MC events for which generated and
reconstructed invariant masses and energies can be compared.
The same procedure as described previously was employed.
When the true and reconstructed \Efivepi --\Minv\ values do not 
agree within their errors, striking peculiarities in the
distribution after the detector simulation are observed.
The expected Gaussian peak, generated by the simulation procedure,
can appear shifted with respect to
its input value (figure\,\ref{fig:tail}a) and sometimes
ambiguities may occur (figure\,\ref{fig:tail}b).

These effects can largely, but not exclusively, be attributed
to pathological track topologies, e.g. hit sharing when tracks 
cross or come very close or when tracks are close to the anode 
plane of the jet chamber. While the shifts as shown in
figure\,\ref{fig:tail}a 
indicate such problematic topologies in a clear way, the quality
cuts described in section\,\ref{Selection} are sufficient
to remove those pathological events in our data sample 
that are sensitive to \Mnutau.
From MC simulation it is estimated that approximately 50\%
of the pathological events are rejected by the cut against
high--momentum tracks in close proximity to a wire plane. 

To assure that this measurement does not deteriorate from such defects,
all candidate events have been individually inspected. None of the
sensitive data events is found to suffer from the discussed biases.
For the two most sensitive events it has been additionally verified 
that the experimental resolution is nearly constant in the
\Efivepi --\Minv\ plane.
Therefore the resolution has been determined for similar events
at several positions
around the data event (figure\,\ref{fig:tail}c,d).
No significant deviation from the resolution of the 
original data event has been found.

\section{Results}
\label{results}
\begin{figure}[htb]
\begin{center}
\resizebox{17cm}{!}{\includegraphics{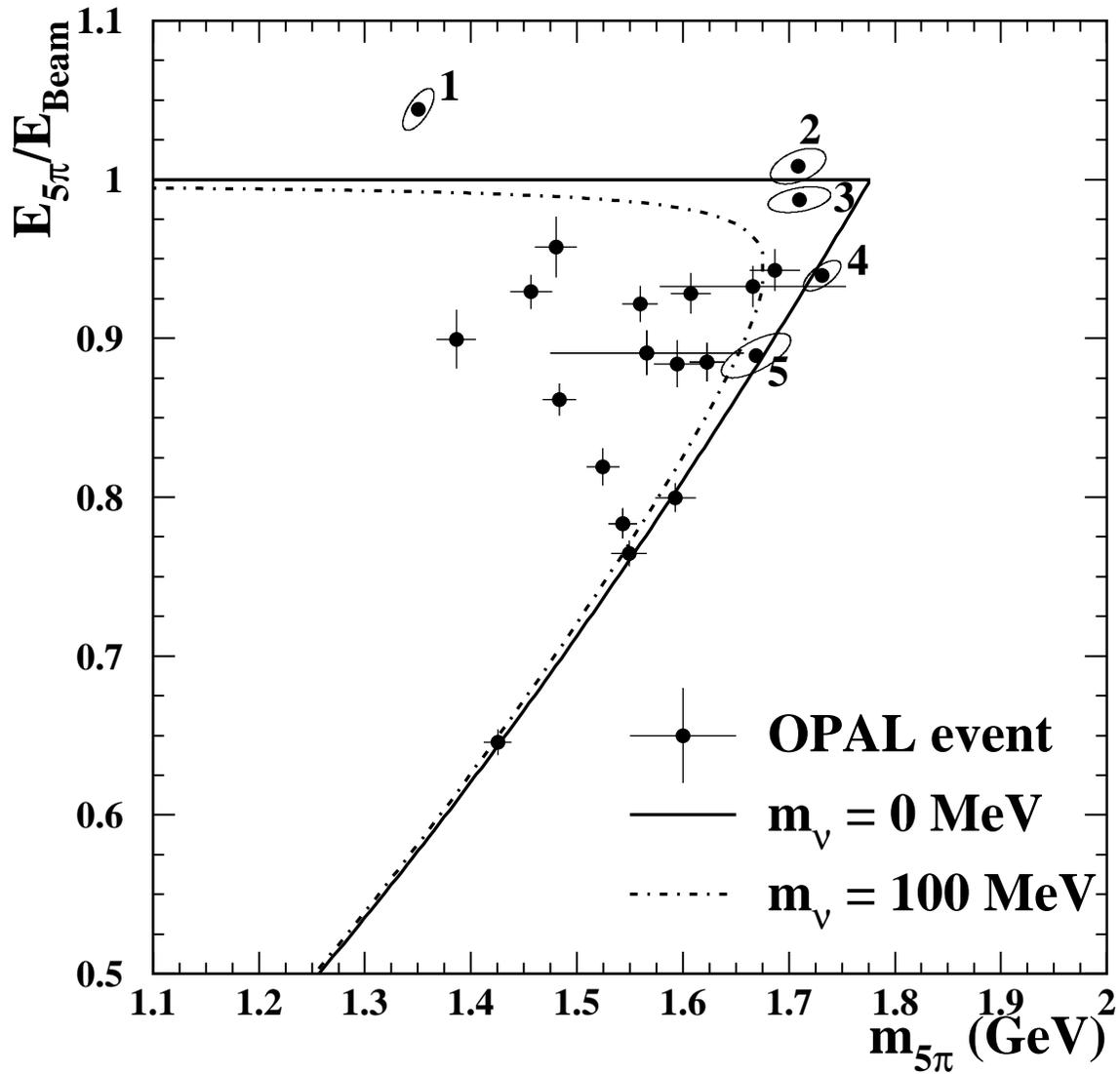}}
\end{center}
\caption{\it Selected events. The sensitive events are numbered (1--5).
The ellipses show the
one standard deviation contour for the 
resolution function. 
The crosses show the errors on \Minv\ and \Efivepi\ for the 
insensitive events calculated 
from the track parameter errors. 
The kinematically allowed regions for massless, 40\,MeV and
100\,MeV \Stau --neutrinos are indicated by the solid,
dashed and dot--dashed lines, respectively.
}
\label{fig:allselected}
\end{figure}

Five sensitive events are retained after
the selection
described in section\,\ref{Selection}.
They are labeled by numbers (figure\,\ref{fig:allselected}).
Three of them (events 2,\thinspace 3,\thinspace 4)
are well reconstructed. 
All tracks of 
the corresponding signal cones are separated sufficiently 
and have at
least 75\% of the maximum number
of possible hits in the jet chamber.

Event 1 contains two high--momentum tracks that
are close to each other
throughout the entire volume of the tracking system.
Thus the reconstruction of this event is likely
to be degraded (see position of this
event in figure\,\ref{fig:allselected}). The effect on the 
limit is small, because the event is located in a less sensitive 
region of the \Efivepi --\Minv\ plane.

In the signal cone of event 5, four track crossings
occur in the jet chamber.
This results in an increased non-Gaussian fraction for the
resolution function in the simulated and reconstructed events 
(section\,\ref{resolution}).
This fact
causes a relatively large error on the invariant mass and energy
by the likelihood fit.

As a result of the background estimation in section\,\ref{background},
the probability that one of the remaining five sensitive events
is background is
1\%.

The upper limit for \Mnutau\ is determined from the 2-dimensional
likelihood technique described in section\,\ref{likelihood}
using the combined likelihood 
of all events which have passed
the selection. This likelihood function is scanned by changing
the assumed true neutrino mass \Mnutau\ in steps of 6\,\MeV.
A third--order polynomial 
multiplied by a Gaussian is used to obtain a functional description
for the likelihood distribution. A 95\% CL 
upper limit of $\Mnutau <$ \TheRawNumber\,\MeV\ is obtained by
integrating the likelihood function over the physical region
of $\Mnutau \geq 0$.
The result is shown in figure\,\ref{fig:combined_limit}a already
including systematic uncertainties as described
in section\,\ref{systematics}.
An alternative (non--Bayesian) approach using the
log--likelihood has been applied as a consistency check,
resulting in an upper limit in agreement with the one
quoted above.

In table\,\ref{tab:minuseins} the impact of each sensitive event  
on the
mass limit is listed.

\begin{table}[htb]
\begin{center}
\begin{tabular}{|c|c|c|}
\hline
Event & Limit variation & Limit from this\\
    &  (\MeV )       &  event alone(\MeV )   \\
\hline
 1    &  +1.8          &   97.7      \\
 2    &  +7.2          &   57.0      \\
 3    &  +3.4          &   66.6      \\
 4    &  +6.1          &   58.0      \\
 5    &  +0.4          &  130.0      \\
\hline 
\end{tabular} 
\caption{ \it Impact of the five sensitive events. The second column
  shows the effect on the mass limit if that
event were to be discarded.
The last column denotes the limit derived from this event alone.}
\label{tab:minuseins}
\end{center}
\end{table}
\section{Systematic errors}
\label{systematics}
The largest
systematic
uncertainty
is from the resolution function, especially from the tail fraction.
For the sensitive events the parameters of the experimental 
resolution are varied by the
average statistical error of the likelihood fit.
The effect on
the limit is small\,(0.5\,\MeV ).

As described in section\,\ref{resolution}
the experimental resolution is determined by the detector
simulation only for sensitive events.
For the remaining events the
errors from the track fit
are used.
The consequences of this on \Minv\ and \Efivepi\ are
estimated by varying these errors by
30\%.
This is the average deviation observed between
the resolution determined by simulation and by calculation
using track parameter errors.
The corresponding variation on the mass limit is 
0.2\,\MeV.

The energy calibration has been checked with 
e$^+$e$^- \rightarrow \mu^+ \mu^-$ events to a level 
of 5$\cdot$10$^{-4}$. The effect on the mass limit due to
this uncertainty is small (0.2\,\MeV).

A contamination of 4 events out of the \Nselected\ selected 
events is expected from the 
decay \Bdfivepizero\ in the lower \Efivepi\ and lower \Minv\ region,
as predicted by MC.
These events do not bias the mass limit
to lower values, because the \Pizero\ is not 
included in the reconstruction. To assess a possible 
impact on the limit, all possible combinations
of four events in the lower \Efivepi-- \Minv\ region
have been successively removed from the event sample
and the limit has been recalculated. 
The shift in the limit 
is +0.5\,\MeV. To be conservative we do not correct 
for this effect.

The structure of the \Sdfivepi\ decay dynamics
only has a very weak effect on the \Mnutau\, limit\,\cite{Aleph:mnutaunew,Cadenas},
negligible as a contribution to the systematics in this analysis.

\begin{table}[htb]
\begin{center}
\begin{tabular}{|c|c|c|}
\hline
 &                Limit variation \\
 \rb{Source}    &  (\MeV )        \\
\hline
 tail fraction description    &  3.5   \\
 resolution function for insensitive events    &  0.2         \\
 resolution function error  &  0.5          \\
 energy calibration         &  0.2 \\
 slope of efficiency    &  0.3          \\
 tau mass               &  0.1    \\
 beam energy    &  0.1          \\
\hline
\hline
 total          & 3.6 \\
\hline 
\end{tabular} 
\caption{ 
\it Systematic effects }
\label{tab:systematics}
\end{center}
\end{table}

As mentioned in section\,\ref{method} the detection efficiency
is assumed to be constant within the kinematic region.
The effect of a possible \Minv -- or \Efivepi --dependent efficiency
has been taken into account by introducing slopes in \Minv\ and 
\Efivepi ,
varying the efficiency by $\mathrm{\pm 25 \%}$. The effect on 
the mass limit was found to be 0.3\,\MeV.

A \Stau\ mass of (1777.0\,$\pm$\,0.3)\,\MeV\
has been used\,\cite{BES:taumass};
its uncertainty leads to a negligible effect on the neutrino
mass limit.

For the beam energy uncertainty, an absolute error of 
4\,\MeV \ and an energy spread of 28\,\MeV\ are assumed
as obtained by the LEP energy working group\,\cite{LEP:energy}.

The systematic
uncertainties are summarized in table\,\ref{tab:systematics}.
All variations of the limit are 
added in quadrature and then added linearly to the mass limit.
The likelihood distribution including all 
systematic uncertainties is shown in
figure\,\ref{fig:combined_limit}a;
it is obtained by scaling the raw likelihood distribution 
(without systematic effects) in \Mnutau\
by the ratio of mass limits obtained with and
without systematic errors. 
This likelihood allows us to combine the results of this
analysis with previous OPAL results on \Mnutau\ including 
systematic errors.
\section{Discussion}
\label{conclusion}

An upper limit for the 
\Stau --neutrino mass has been derived using the \Stau\ decay 
mode \Sdfivepi. 
Including
systematic uncertainties,
the upper limit \\
\begin{center}
\Mnutau \smaller\,\TheFinalNumber\,\MeV\  is obtained at 95\% confidence level.\\
\end{center}
\vspace*{5mm}

This result is based on a data sample that is five times larger than
the result previously published by OPAL for this channel\,\cite{Opal:first}, and leads to a
significant improvement of the limit.

\begin{figure}[htb]
\begin{center}
\resizebox{17cm}{!}{\includegraphics{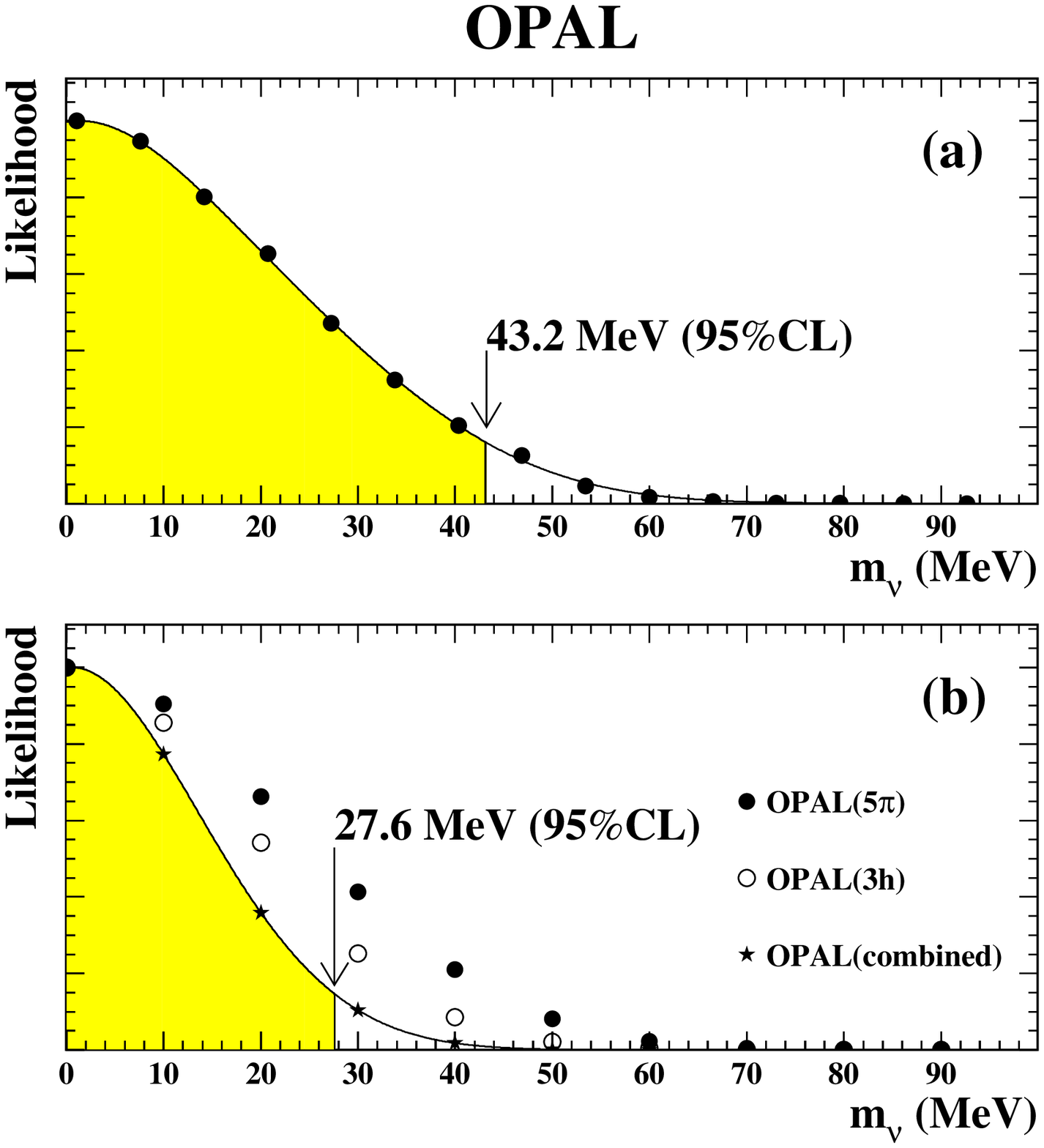}}
\end{center}
\caption{\it Likelihood versus \Mnutau\ for (a) this
analysis using \Sdfivepi\ and  (b) combined with the OPAL
\Makotodecay\ analysis. The effects of systematic uncertainties
are included.}
\label{fig:combined_limit}
\end{figure}

The combination of this measurement with the previously
published OPAL analysis
using \Makotodecay\ decays\,\cite{Opal:makoto} is obtained by multiplying
the respective likelihood curves including the systematic
uncertainties.
The result is shown in figure\,\ref{fig:combined_limit}b.
From the combined new likelihood curve,
the upper limit\\
\begin{center}
 \Mnutau \smaller\,\TheCombinedNumber\,\MeV\  is obtained at 95\% confidence level.\\
\end{center}
\vspace*{5mm}

Similar limits of 31\,\MeV\ and 30\,\MeV\ have been also obtained
by the ARGUS\,\cite{Argus:limit} and CLEO\,\cite{Cleo:newmnutau}
experiments, respectively.
Recently, a new upper limit of 18\,\MeV\ has been determined 
by the ALEPH Collaboration\,\cite{Aleph:mnutaunew},
also by using the results from three- and five-prong tau decays.
Thus, a tau--neutrino mass of
less than 30\,\MeV\ is well established and confirmed by
several experiments.

\newpage
\noindent
{\large\bf Acknowledgements:}
\par
\noindent
We particularly wish to thank the SL Division for the efficient operation
of the LEP accelerator at all energies
 and for their continuing close cooperation with
our experimental group.  We thank our colleagues from CEA, DAPNIA/SPP,
CE-Saclay for their efforts over the years on the time-of-flight and trigger
systems which we continue to use.  In addition to the support staff at our own
institutions we are pleased to acknowledge the  \\
Department of Energy, USA, \\
National Science Foundation, USA, \\
Particle Physics and Astronomy Research Council, UK, \\
Natural Sciences and Engineering Research Council, Canada, \\
Israel Science Foundation, administered by the Israel
Academy of Science and Humanities, \\
Minerva Gesellschaft, \\
Benoziyo Center for High Energy Physics,\\
Japanese Ministry of Education, Science and Culture (the
Monbusho) and a grant under the Monbusho International
Science Research Program,\\
German Israeli Bi-national Science Foundation (GIF), \\
Bundesministerium f{\"u}r Bildung, Wissenschaft,
Forschung und Technologie, Germany, \\
National Research Council of Canada, \\
Research Corporation, USA,\\
Hungarian Foundation for Scientific Research, OTKA T-016660, 
T023793 and OTKA F-023259.\\
\clearpage

\clearpage
\clearpage
\clearpage
\clearpage
 

\begin{thebibliography}{10}

\bibitem{theory:darkmatter}
J.~Primack, D.~Seckel, and B.~Sadoulet,
\newblock Ann.Rev.Nucl.Part.Sci. {\bf 38} (1988) 751.

\bibitem{Theory:seesaw}
M.~Gell-Mann, P.~Ramond, and R.~Slansky,
\newblock {\em Supergravity},
\newblock ed. by D. Freedman et al., North Holland, 1979.

\bibitem{Theory:gravlimit}
R.~Cowsik and J.~McClelland,
\newblock Phys. Rev. Lett. {\bf 29} (1972) 669.

\bibitem{Theory:bigbang}
M.~Kawasaki et~al.,
\newblock Nucl.Phys. {\bf B419} (1994) 105.

\bibitem{Opal:first}
{OPAL Collaboration, R. Akers} et~al.,
\newblock Zeit. f. Physik {\bf C65} (1995) 183.

\bibitem{Opal:detector1}
{OPAL Collaboration, K. Ahmet} et~al.,
\newblock Nucl. Instrum. Meth. {\bf A305} (1991) 275.

\bibitem{Opal:Silicon1}
P.~Allport et~al.,
\newblock Nucl. Instrum. Meth. {\bf A324} (1993) 34.

\bibitem{Opal:Silicon2}
P.~Allport et~al.,
\newblock Nucl. Instrum. Meth. {\bf A346} (1994) 476.

\bibitem{Opal:cjperformance}
O.~Biebel et~al.,
\newblock Nucl. Inst. Meth. {\bf A323} (1992) 169.

\bibitem{Opal:hausch1}
M.~Hauschild et~al.,
\newblock Nucl. Inst. Meth. {\bf A314} (1992) 74.

\bibitem{Opal:hausch2}
M.~Hauschild,
\newblock Nucl. Inst. Meth. {\bf A379} (1996) 436.

\bibitem{Generator:tauola-24}
R.~Decker, S.~Jadach, J.~K{\"u}hn, and Z.~W\c{a}s,
\newblock Comput. Phys. Commun. {\bf 76} (1993) 361.

\bibitem{Generator:koralz-40}
S.~Jadach, B.~Ward, and Z.~W\c{a}s,
\newblock Comput. Phys. Commun. {\bf 79} (1994) 503.

\bibitem{Generator:jetset74}
T.~Sj\"{o}strand,
\newblock Comp.~Phys.~Comm. {\bf 82} (1994) 74.

\bibitem{Generator:fermisv}
J.~Hilgart, R.~Kleiss, and F.~{Le Diberder},
\newblock Comp.~Phys.~Comm. {\bf 75} (1993) 191.

\bibitem{Opal:gopal}
J.~Allison et~al.,
\newblock Nucl. Inst. Meth. {\bf A317} (1992) 47.

\bibitem{Opal:cone}
{OPAL Collaboration, G. Alexander} et~al.,
\newblock Zeit. f. Physik {\bf C52} (1991) 175.

\bibitem{Opal:h3paper}
{OPAL Collaboration, R. Akers} et~al.,
\newblock Zeit. f. Physik {\bf C68} (1995) 555.

\bibitem{Aleph:mnutaunew}
{ALEPH Collaboration, R. Barate} et~al.,
\newblock Eur. Phys. J. {\bf C2} (1998) 395.

\bibitem{Cadenas}
J.~G\'omez-Cadenas,
\newblock Sensitivity of Future High--Luminosity e$^+$e$^-$ colliders to a
  Massive $\tau$ Neutrino,
\newblock in {\em Third workshop on the tau--charm factory}, pages 97--140,
  1993.

\bibitem{BES:taumass}
{BES Collaboration, J. Z. Bai} et~al.,
\newblock Phys. Rev. {\bf D53} (1996) 20.

\bibitem{LEP:energy}
\mbox{LEP} energy~working group,
\newblock Calibration of centre-of-mass energies at LEP1 for precise
  measurements of Z properties, 1998,
\newblock CERN-EP/98-40; submitted to European Physical Journal C.

\bibitem{Opal:makoto}
{OPAL Collaboration, G. Alexander} et~al.,
\newblock Zeit. f. Physik {\bf C72} (1996) 231.

\bibitem{Argus:limit}
{ARGUS Collaboration, H. Albrecht} et~al.,
\newblock Phys. Lett. {\bf B292} (1992) 221.

\bibitem{Cleo:newmnutau}
{CLEO Collaboration, R. Ammar} et~al.,
\newblock A limit on the mass for the tau neutrino, 1998,
\newblock CLNS-98-1551.

\end{thebibliography}
\end{document}